\documentclass{emulateapj}
\slugcomment{{\em}}
\usepackage{epsf}

\begin{document}
\title{Modeling The Star Forming Universe at z=2: Impact of Cold Accretion Flows}

\author{S. Khochfar$^{1}$,  Joseph Silk$^2$}
 
\affil{$^1$ Max-Planck-Institute for Extraterretrial Physics, Giessenbachstr., D-85748, Garching, Germany \\ 
$^2$ Department of Physics, Denys Wilkinson Bldg., University
of Oxford, Keble Road, Oxford, OX1 3RH, UK }

\begin{abstract}
We present results of a semi-analytic model (SAM) that includes cold accretion and a porosity-based prescription for star formation. We can recover the puzzling observational results of low $V/\sigma$ seen in various massive disk or disk-like galaxies, if we allow 18 $\%$ of the accretion energy from cold flows to drive turbulence in gaseous disks at $z=2$. The increase of gas mass through cold flows is by itself not sufficient to increase the star formation rate sufficiently to recover the number density of $\dot{M}_*>120$ M$_{\odot}$ yr$^{-1}$ galaxies in our model.  In addition, it is necessary to increase the star formation efficiency. This can be achieved naturally in the porosity model, where star formation efficiency scales $\propto \sigma$, which scales as cloud velocity dispersion. As cold accretion is the main driver for gas velocity dispersion in our model, star formation efficiency parallels cold accretion rates, and allows  fast conversion into stars. At $z\sim 2$, we find  a space density $~10^{-4}$ Mpc$^{-3}$ in star-forming galaxies with $\dot{M}_*>120$ M$_{\odot}$ yr$^{-1}$, in better agreement than earlier estimates form SAMs. However, the fundamental relation between $\dot{M}_*$ and $M_*$ is still offset from the observed relation, indicating the need for possibly  more efficient star formation at high z perhaps associated with a role for AGN triggering. 
\end{abstract}

\section{Introduction}
Multiple lines of evidence suggest that galaxy formation is taking place with a varying degree of efficiency over cosmic time. The overall star formation density in the universe declines at $ z\le 2$ \citep[e.g.][and references therein]{2004ApJ...615..209H}, a feature seen also in the normalization of the relation between stellar mass and star formation rate \citep{2006ApJ...637..727C,2007ApJ...670..156D,2007ApJ...660L..43N}. Further support for the notion of rapid early formation of galaxies at $z \geq 2$, especially massive ones, comes from high $\alpha/$Fe ratios \citep{2005ApJ...621..673T}, early metal enrichment \citep{2008A&A...488..463M}, growth of super-massive black holes \citep{2005A&A...441..417H} and the assembly rate of massive galaxies \citep{2008ApJ...675..234P}. The redshift range $1.4 \leq z \leq 2.5$ has been targeted by various surveys combining broadband color selection techniques \citep[e.g.][(D07)]{2004ApJ...617..746D,2004ApJ...604..534S,2007ApJ...670..156D} and recently by high-resolution H$\alpha$ integral field spectroscopy \citep{2006ApJ...645.1062F}. The latter reveals massive $ \sim 10^{11}$M$_{\odot}$, rotating disk galaxies with star formation rates of $\sim 140$ M$_{\odot}$ yr$^{-1}$ and velocity dispersions of $\sim 60$ km s$^{-1}$   \citep[e.g.][]{2006Natur.442..786G}. The origin of the high star formation rates and the large velocity dispersion in these galaxies is not presently understood.   

On the theoretical side, detailed comparisons between observations at $z=2$ and N-body-SPH simulations or semi-analytical models (SAM) show  discrepancies. Although the overall slope of the $M_*-$SFR is well recovered in the models, the normalization is significantly smaller \citep[D07;][see however Dekel et al. 2008]{2008MNRAS.385..147D}. 

 Theory and numerical simulations show that galaxies accrete cold gas 
in two modes. At early times and in small dark matter halos, most of the gas never shock heats to the  virial temperature and is accreted on a free fall time scale \citep{2003MNRAS.345..349B,2005MNRAS.363....2K,2006MNRAS.368....2D,2008MNRAS.390.1326O,2009MNRAS.395..160K}. The general condition for a stable extended shock to be maintained is that the post-shock gas remains highly pressurized. In practice, this means that the loss of energy by radiative cooling needs to be at least compensated by the compression of post-shock infall material \citep{2006MNRAS.368....2D}. This condition leads to a critical halo mass of M$_{shock} \sim 10^{11.6}$ M$_{\odot}$ \citep{2008MNRAS.390.1326O} for shock heating of gas and hence a transition from fast and efficient "cold accretion" onto the galaxy to  a more  moderate "hot accretion " mode set by the radiative cooling time of the gas.  In an earlier study \citet{2008ApJ...688..789G} showed that the number density and accretion rates of dark matter halos  suffice to provide enough high star forming galaxies, assuming that the newly accreted dark matter provides the universal fraction of baryons, and that these baryons are all transformed into stars. Alternatively, using a hydrodynamical simulation, \citet{2009Natur.457..451D} showed that the gas accretion rates are comparable to the observed star formation rates thus requiring a very efficient transformation of accreted gas into stars.

In this Letter,  we aim at addressing the interlinked  questions  of high star formation rates and velocity dispersion observed in galaxies at $z=2$. Our work expands on the results previously presented by \citet{2009Natur.457..451D} and \citet{2008ApJ...688..789G}.  In particular, we model the effect of cold accretion on driving turbulence in the interstellar medium (ISM) and employ a porosity description for the ISM and star formation within it. The latter hypothesis is able to provide high star formation efficiencies in the context of a physically motivated model.

\section{Model}
In the following, we briefly outline our fiducial semi-analytic model \citep[][and references therein]{2005MNRAS.359.1379K,2006MNRAS.370..902K} and the changes we have introduced. We generate merger trees for dark matter halos 
using a Monte-Carlo approach based on the extended Press-Schechter formalism 
\citep{1999MNRAS.305....1S} down to a mass resolution of $M_{min}=5 \times 10^{9}$. Once a tree reaches $M_{min}$, we start moving 
the tree forward in time including physical processes 
associated with the baryons within each dark matter halo that include
gas cooling, star-formation, supernova feedback, reionization and merging 
of galaxies on a dynamical friction time-scale. In this Letter, we will omit 
including detailed prescriptions for AGN-feedback \citep[e.g.][]{2006Natur.442..888S} or environmental effects \citep{2008ApJ...680...54K}.

 In our modeling approach, we try to include the cold and hot accretion modes by first assigning the fraction $ f_{cold}$ of gas within the halo, that is in a cold accretion flow. If $M_{halo,g}$ is the mass of gas within the halo, then we can readily calculate the amount of gas in the hot phase as $(1-f_{cold}) M_{halo,g}$. We use an approximate fit to the fraction of cold accreted material as shown in Fig. 3 of \citet{2009MNRAS.395..160K} and the bottom part of Fig. 4 in \citet{2008MNRAS.390.1326O}. Please note that the  results of  \citet{2009MNRAS.395..160K} indicate a mild evolution with redshift in this relation, which we for the sake of simplicity neglect in this study. Throughout this work we will use $f_{cold}=-1.6 + 0.89 \log(M_{vir})-0.06(\log(M_{vir}))^2$, with a lower and upper floor of $0$ and $1$ for halos greater $\sim 10^{12.7}$ M$_{\odot}$ and less $ \sim 10^{10.74}$ M$_{\odot}$, respectively. In a second step, we then calculate the cold and hot mode accretion onto the galaxy via $\dot{M}_{cool,c}=f_{cold}M_{halo,g}/t_{dyn}$ and $\dot{M}_{cool,h}=0.5 (1-f_{cold})(M_{hot} r_{cool})/(t_{dyn} R_{vir}) $, respectively. Part of the underlying assumption is that cold accretion flows will reach the galaxy on a halo dynamical time $t_{dyn}$, and that the hot accretion rate will be governed by the radiatively cooling time using recipes laid out as in e.g. \citet{2001MNRAS.328..726S}. We find in general that cold accretion clearly dominates over hot accretion in agreement with results of numerical simulations \citep[e.g.][]{2009Natur.457..451D,2009MNRAS.395..160K}. The accretion rate of gas onto the disk of the central galaxy is then approximated as $\dot{M}_{cold}=\dot{M}_{cool,c}+\dot{M}_{cool,h}$. This choice of accretion rate is the simplest possibility and shows results in agreement with those from hydrodynamical simulations presented by \citet{2009Natur.457..451D}.

Star formation is modeled following the Schmidt-Kennicutt-law $ \dot{M}_*= \alpha M_{cold}/t_{dyn,disc}$  \citep{1998ApJ...498..541K} with a porosity-based prescription for the multiphase structure of the ISM \citep{2001MNRAS.324..313S}. Momentum balance  yields a star formation efficiency 
$\alpha \approx 0.02 (\sigma_g/10$ km s$^{-1})$ $(v_{cool}/400$ km s$^{-1})$   $(m_{SN}/ 150$ M$_{\odot})$ $(10^{51}$ erg$/E_{SN})$ \citep{2005MNRAS.364.1337S}. 
 The $\sigma_g$ dependence arises because of the enhancement in star formation efficiency via momentum input from supernova remnants  with increasing ambient turbulent pressure. Note that gas accretion also drives gas turbulence, as in our model. Momentum input is only effective once the remnant expansion speed has slowed to below $v_{cool}.$
Here $\sigma_g$ is the gas velocity dispersion, $m_{SN}$ is the IMF-dependent mass in stars formed for each type II supernovae, $E_{SN}$ the energy released by each supernova and  $v_{cool}$ the shell velocity at which strong radiative energy losses set in. We set $v_{cool}= 400$ km s$^{-1}$ and use a Chabrier IMF with $m_{SN}=150$ M$_{\odot}$ throughout this paper. To calculate the gas velocity dispersion in the disk, we assume equilibrium between the rate at which energy is dissipated in the disk and energy added due to kinetic energy of infalling gas and supernovae.  Including the $\sigma$ dependence of the star formation rate predicted by the porosity model, this leads to $\sigma=\beta+(\beta^2+ \eta \dot{M}_{cool,c}t_{dyn,disc}/M_{cold}V^2)^{1/2}$. The free parameter $\eta$ allows to adjust for our ignorance with respect to the actual fraction of the kinetic energy being used to drive turbulence and the relation between the dissipation time-scale and dynamical time of the disk. $\eta$ is set to match the observed relation between $V$ and $V/\sigma$ of galaxies at $z \sim 2$. The constant $\beta=10/2$ km s$^{-1}$ is set to ensure that in the limiting case of no accretion, the local observed velocity dispersion in disks is recovered. Simulations show that cold accretion comes in along filaments on almost radially orbits before hitting the galactic disk \citep{2008MNRAS.390.1326O}, thus being able to contribute a large fraction of the gas  kinetic energy to the heating of the gaseous disk,  in contrast to 
hot-mode accreted material that settles more gradually onto the disk.
 
In this study, we simulate a volume of $10^6$ Mpc$^3$, and use the following set of cosmological parameters derived from a combination of the 5-year WMAP data with Type Ia supernovae and measurements of baryon acoustic oscillations \citep{2008arXiv0803.0547K}:
$\Omega_0=0.28$, $\Omega_{\Lambda}=0.72$, $\Omega_b/\Omega_0=0.16$, 
$\sigma_8=0.8$ and $h=0.7$. 

\section{Low $V/\sigma$ Disk Galaxies}
We start by investigating the increase in gas velocity dispersion in gaseous disks due to cold accretion. Fig. 1 shows the correlation between $V/\sigma$ and $V$. In general only $18\%$ $(\eta=0.18)$ of the accretion energy is sufficient to drive turbulence that agrees with the range of observed values \citep{2006Natur.442..786G,2008ApJ...687...59G,2008Natur.455..775S}. It is interesting to note that this value is in agreement with recent theoretical predictions by \cite{2009arXiv0901.2458D}.
Cold accretion-driven turbulence in galaxies living in small dark matter halos with $V_{c} < 100$ km s$^{-1}$ results in velocity dispersions similar to or larger than the rotational velocity,  mainly because of the high accretion rates and the low gas masses in the disks to absorb the associated accretion energy. These galaxies are dynamically hot system that are pressure-supported against gravitational collapse in contrast to classical cold disks. More massive galaxies with larger $V$ typically live in massive halos which only have small fractions of cold accretion via filaments and massive gaseous disks that are not easily heated.  The steep transition from low to high $V/\sigma$ reflects the steep decline of cold accretion flows as a function of dark halo mass that we have adopted based on simulations.
The scatter seen in the $V/\sigma - V$ plot is largely due to the varying accretion histories of individual halos.    
\begin{figure}
  \plotone{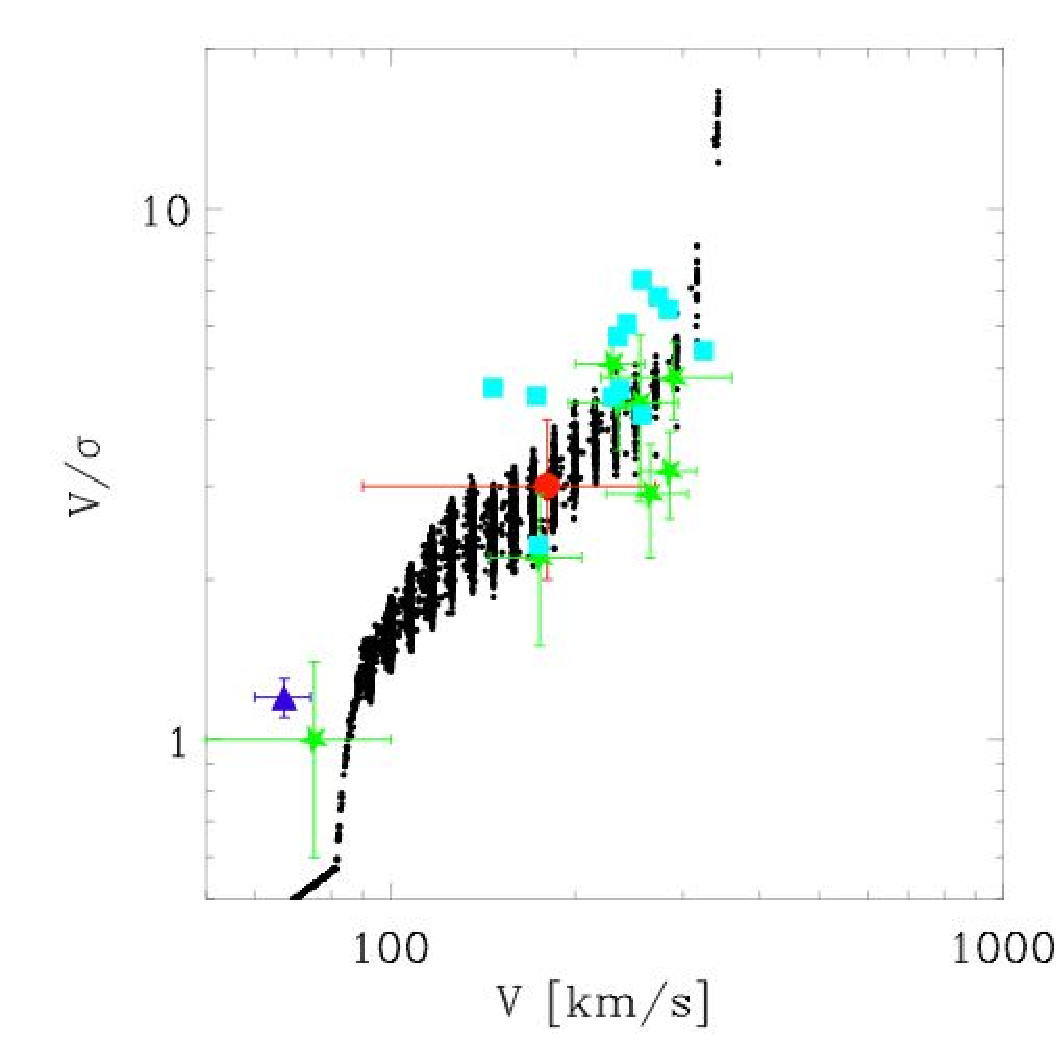}\label{f1}
  \figcaption{The relation between $V/\sigma$ and $V$ for modeled galaxies (small filled circles). The observations are from \citet{2006Natur.442..786G} (large filled circle), \citet{2008ApJ...687...59G} (filled stars), \citet{2008Natur.455..775S} (filled triangle) and Cresci (2008, ApJ submitted) (filled squares). We here assume that $18\%$ of the accretion energy from cold flows is used to drive turbulence in the disk. Note that mergers are excluded as well as satellite galaxies that by construction do not accrete cold gas anymore.} 
\end{figure}

\section{Star Formation Rates and Mass Function}
 The inclusion of explicit cold accretion in the SAM increases the overall accretion rate onto model galaxies. This might suggest that the star formation rates will increase as well. However, using a constant star formation efficiency normalized to the locally observed value of $\alpha=0.02$ the number density of star forming galaxies with $\dot{M}_*> 120$ M$_{\odot}$ yr$^{-1}$ falls short by almost an order of magnitude in agreement with previous reports by D07 and \citep{2008MNRAS.385..147D}.
Fueling of  the galaxies with gas is not the limiting factor for high star formation rates. Rather, it is the conversion of gas into stars over a short time-scale. The porosity-based model for star formation predicts that star formation efficiency scales as $\alpha \propto \sigma$, thus providing  a natural way to increase the star formation rate  with redshift in parallel with the cold gas accretion rate. 

In Fig. 2, we present the SFR-$M_*$ relation for central galaxies at $z=2$. The model galaxies cover most of the observed range. However, we find that the slope is slightly tilted. Mostly massive model galaxies have too low star formation rates. Inspection of the D07 data reveals that the statistics at the massive  end is rather poor and that a possible flattening at high masses as seen in our models cannot be ruled out. 
It is interesting to note that the modeled SFR$-M_*$ relation at $z=2.5$ lies perfectly on the region covered by the observed galaxies. The number density of star-forming galaxies with $\dot{M}_*> 120$ M$_{\odot}$ yr$^{-1}$ at $z=2$ is $~ 10^{-4}$ Mpc$^{-3}$ and within a factor two of the observed density of $1.6 \times 10^{-4}$ in D07. Again our estimates for $z=2.5$ are in excellent agreement with the results of D07 at $z=2$. The number density of star-forming galaxies increases by a factor two to $z=3.5$ and then starts to drop again (see Fig. 3).  \citet{2009Natur.457..451D} find in their simulation a similar increase in the number density of galaxies accreting gas with $120$ M$_{\odot}$ yr$^{-1}$ going from $z=2$ to $z=3-4$. However, their overall number densities are larger by a factor $\sim 2$ or more. Also do they not find a peak around $z\sim 3$ as we do, but a steady decline in the number density of systems accreting with high rates.   
\begin{figure}
  \plotone{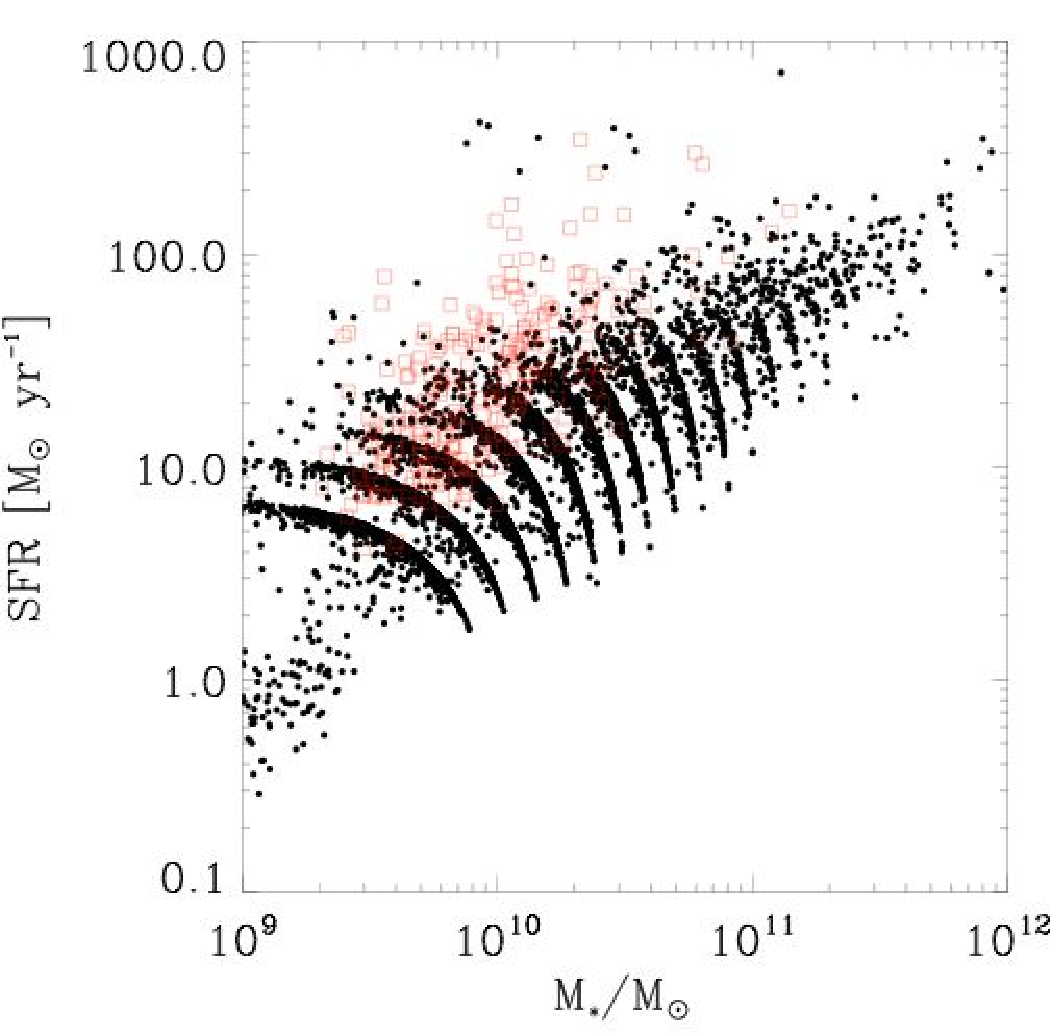}\label{f2}
  \figcaption{Relation between star formation rates and stellar mass for central galaxies. The open squares show the D07 data. The pattern in the model data is due to the binning of halo masses along the halo mass function at the beginning of the merging tree at $z=2$.}
\end{figure}

     As we are not generating light cones to compare with the observations, but use redshift slices, we need to ask the question whether the ULIRG sub-sample of D07 is biased toward $z>2$ within the overall sample which spans the range $ 1.4 \leq z \leq 2.5$. Inspecting their Fig. 6, we find that galaxies with L$_{IR}> 10^{12}$ L$_{\odot}$ span a wide range in redshift between $z=1.5$ and $z=3$ with brighter galaxies preferably being at $z>2$, suggesting that we are underestimating number densities because we neglect the $z>2$ contribution. Taking this into account, the difference with  D07 should be even less than a factor of  two.   

Finally, we turn to the overall mass function of galaxies at $z=2$, shown in Fig. 4. One might expect that due to the enhancement of early star formation in the revised model, basic properties like the mass function will change significantly. However, good agreement is found between the modeled mass function and those based on Spitzer observations by \citet{2005ApJ...630...82P}  and a recent study by \citet{2008arXiv0811.1773M}. The GOODS-MUSIC results of \citet{2006A&A...459..745F} are a bit lower, but agree within their error margin. 

\section{Conclusions}
We have investigated a possible scenario for the formation of galaxies at high redshifts that is driven mainly by cold accretion flows. Massive galaxies at $z=2$ live in halos that are at the intersections of filaments that form the cosmic web. As shown in numerical simulations, these filaments provide a steady stream of dense cold gas that can reach the galaxy on a free-fall time scale and fuel ongoing star formation. Not only do these streams provide cold gas, but they can also drive turbulence in the gaseous disk by depositing some fraction of the accretion energy into the ISM. We find that between $10\%-20\%$ of this accretion energy is sufficient to increase the gas velocity dispersion to observed values. In particular we choose $\eta=0.18$, which still leaves 
accretion driven turbulence dominant with respect to supernovae driven one.  A natural consequence of cold accretion driving turbulence is that it becomes less important in massive halos  $> 10^{12.5}$ $M_{\odot}$ and at low redshifts, \begin{figure}
  \plotone{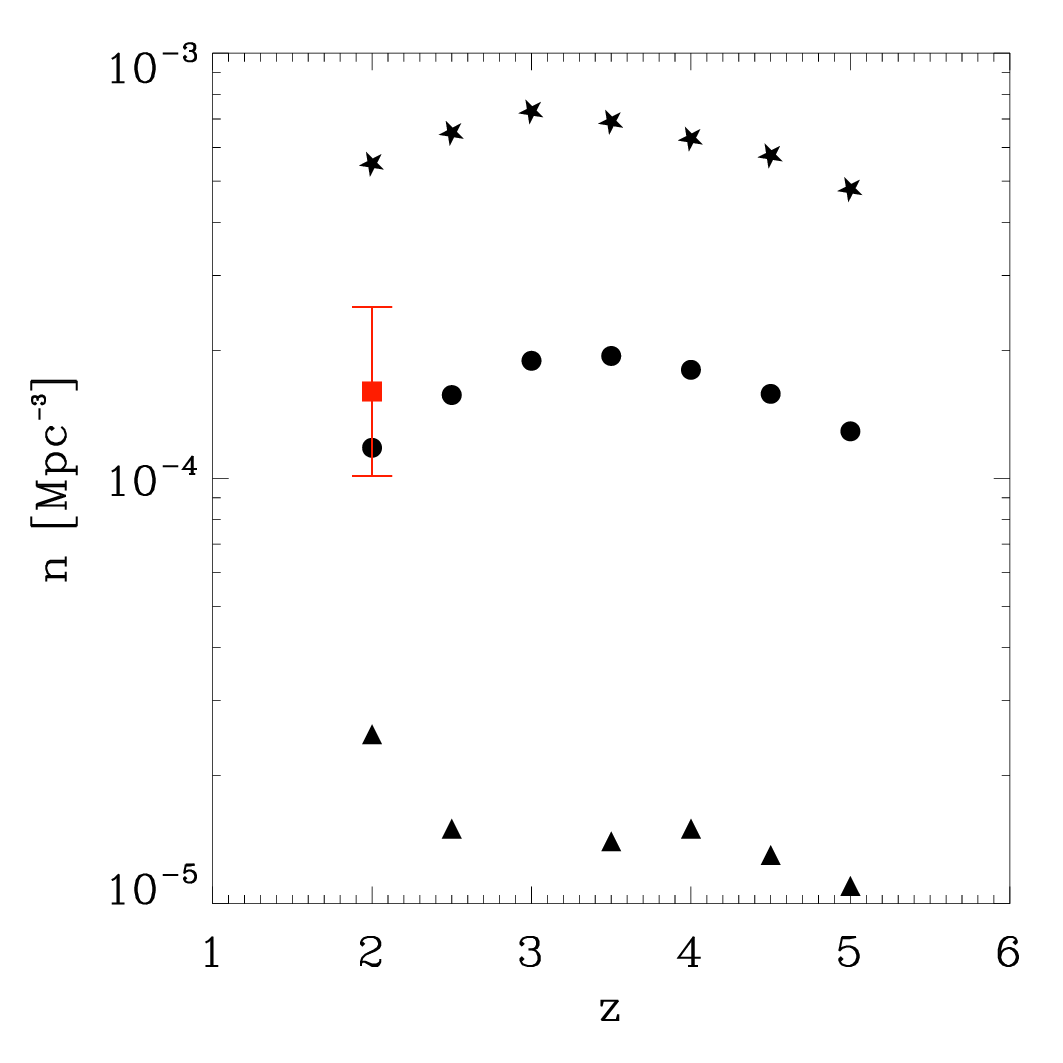}\label{f3}
  \figcaption{Number density of star-forming galaxies with SFR  $>60$ M$_{\odot}$ yr$^{-1}$ (stars), $ > 120$ M$_{\odot}$ yr$^{-1}$ (circles) and  $> 300$ M$_{\odot}$ yr$^{-1}$ (triangles) . The square is the data of D07 for galaxies with  SFR $> 120$ M$_{\odot}$ yr$^{-1}$.}
\end{figure}
where the fraction of cold mode accretion drops and $V/\sigma$ increases. Low redshift observations find velocity dispersions of 10 km s$^{-1}$, much lower than those at $z=2$. This can be understood as a natural consequence of the steady demise of cold accretion with redshift in halos that host disk galaxies today. The host halos of these galaxies live no longer at the intersection of filaments, but in the filaments itself. Those galaxies living in halos above the critical dark matter mass for stable shocks \citep{2006MNRAS.368....2D} will no longer accrete from cold flows, but from hot halo gas, if at all, that radiatively cools and the main driver of turbulence will be supernovae. Such  a picture does fit in nicely with observation of low redshift disk galaxies, which have much larger $V/\sigma$ than their high-$z$ counter parts \citep{2007ApJ...660L..35K}.    

Only increasing the amount of available fuel due to cold accretion is not enough to increase the number density of galaxies with large star formation rates. It is necessary to also increase the star formation efficiency at the same time. We \begin{figure}
  \plotone{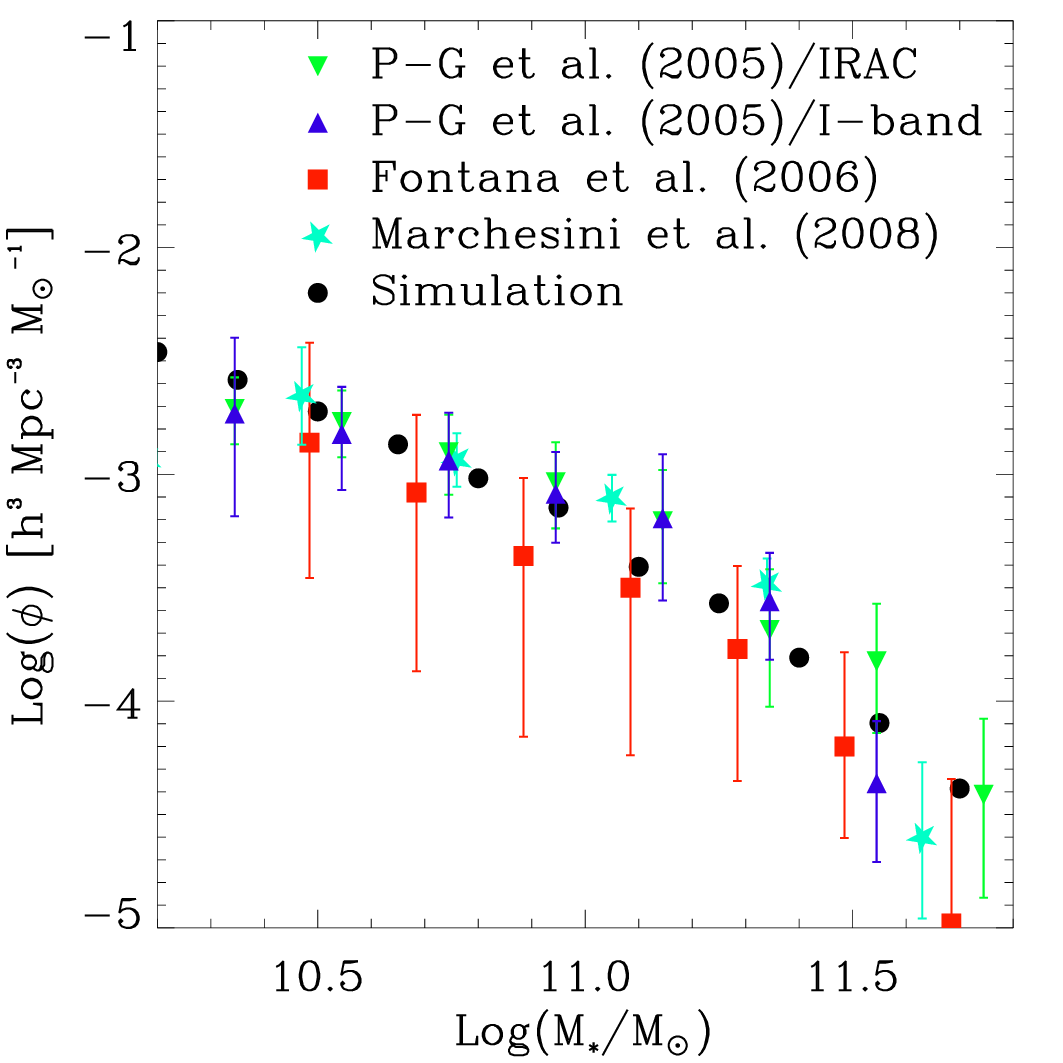}\label{f4}
  \figcaption{Galaxy mass functions at $z=2$. Filled circles show results including cold accretion and the porosity based model for star formation.    }
\end{figure}
tested an analytic model for a multiphase ISM that is based on a porosity 
prescription, that has been shown to capture some of the relevant physical mechanism related to star formation seen in high-resolution simulations \citep{2005MNRAS.356..737S}. The star formation efficiency in this model is 
$\alpha \propto \sigma$, resulting in an increasing efficiency that parallels the gas supply by cold accretion and allows to convert it into stars on a very short 
time-scale. With this model we find improved agreement with the observed number density of star-forming galaxies.   However our predicted star formation rates are lower than in the observed $z\sim 2$ sample especially for the more massive galaxies. We have two comments on this point. Firstly,  the data are most likely biased to high star formation rates. Secondly, galaxies with high star-forming rates at high redshift commonly show  weak AGN activity (see for example  a recent study of a sample of $z\sim 1.9 $ ULIRGs \citep{2009huang} ). This most likely is the tip of the AGN iceberg  because of  possible obscured AGN. AGN outflows could trigger as well as play the more accepted  role of quenching star formation at high redshift.
It remains to be investigated how well the agreement at different redshifts is, and we plan on addressing this in a future paper.
\\

We would like to thank the referee Avishai Dekel for his very helpful comments that helped improve the manuscript, Andreas Burkert and the SINS team for useful discussions, and Emanuele Daddi as well as Giovanni Cresci for kindly providing their observational data.

\end{document}